\documentclass[aps,pra,reprint, showpacs, superscriptaddress]{revtex4-1}

\usepackage{epsfig,graphics,graphicx}
\usepackage{dcolumn}
\usepackage{bm}
\usepackage{color}
\usepackage{amsmath,amssymb,amsthm}
\usepackage{centernot}
\usepackage{bbold}
\usepackage{color,soul}
\usepackage{mathtools}
\usepackage{fullpage}
\usepackage{units}
\usepackage[usenames,dvipsnames]{xcolor}
\usepackage{braket}
\usepackage{indentfirst}
\usepackage{lipsum}
\usepackage[export]{adjustbox}
\usepackage{enumerate}

\usepackage[utf8]{inputenc}
\usepackage[T1]{fontenc}

\usepackage{dsfont}

\usepackage{mathptmx}
\usepackage{amssymb}
\usepackage{amsmath}
\usepackage{latexsym}
\usepackage{amsfonts}
\usepackage{float}

\usepackage{hyperref}
\hypersetup{pdfpagemode=UseNone}

\newtheorem{theorem}{Theorem}
\newtheorem{definition}[theorem]{Definition}

\newtheorem{lemma}[theorem]{Lemma}
\newtheorem{proposition}[theorem]{Proposition}

\newcounter{rem}
\def\>{\rangle}
\def\<{\langle}
\newcommand{\proj}[1]{| #1 \rangle\! \langle #1 |}

\renewcommand{\rho}{\varrho}
\newcommand{\idty}{\mathds{1}}

\def\textbf#1{{\bf #1}}

\def\beq{\begin{equation}}
\def\eeq{\end{equation}}
\def\beqa{\begin{eqnarray}}
\def\eeqa{\end{eqnarray}}
\def\eea{\end{array}}
\def\bea{\begin{array}}
\newcommand{\bei}{\begin{itemize}}
\newcommand{\eei}{\end{itemize}}
\newcommand{\bee}{\begin{enumerate}}
\newcommand{\eee}{\end{enumerate}}
\def\bep{\begin{proposition}}
\def\eep{\end{proposition}}
\def\bel{\begin{lemma}}
\def\eel{\end{lemma}}
\def\bet{\begin{theorem}}
\def\eet{\end{theorem}}
\def\bed{\begin{definition}}
\def\eed{\end{definition}}

\begin{document}

\title{Decay of quantumness in a measurement process: action of a coarse graining channel}

\author{Gabriel Dias Carvalho}
\email{gabrieldiascarvalho@outlook.com}
\affiliation{Departamento de Física, Universidade Federal de Pernambuco, Recife, Pernambuco, CEP 50670-901l}
\affiliation{Centro Brasileiro de Pesquisas F\'isicas, Rio de Janeiro, Rio de Janeiro, CEP 22290-180}

\author{Pedro Silva Correia}
\email{pedrosc8@cbpf.br}
\affiliation{Centro Brasileiro de Pesquisas F\'isicas, Rio de Janeiro, Rio de Janeiro, CEP 22290-180}

\date{\today}


\begin{abstract}

A model of a quantum measurement process is presented: a system consisting of a qubit in a superposition interacts with a measuring apparatus consisting of a $N$ qubit state. Looking at the emerging, effective description of the apparatus given by the action of a coarse graining channel, we have been able to recover information about the superposition coefficients of the system. We have also been able to visualize the death of quantum correlations between system and apparatus and the death of quantum coherences in the apparatus' effective state, in the limit of a strong coarse graining action. A situation akin to decoherence, although it is not necessary to evoke any interaction with the surrounding environment.

\end{abstract}


\maketitle



\section{Introduction}

Currently, nature's best description is given by quantum mechanics. According to quantum theory, to every system we assign a quantum state $\psi$, and its evolution is dictated by Schrödinger's equation \cite{cohen77}. As Schrödinger himself realized, if we take this postulate to our macroscopic and everyday world, we quickly run into paradoxical situations -- for example the possibility of an alive-and-dead cat \cite{schrodinger35}. Not only we do not observe quantum effects on macroscopic systems, but also we do not employ the full quantum description for such systems. In fact, our everyday life experiences heavily rely on effective (macroscopic) descriptions which are far less complex than their underlying quantum characterization.

The quantum-classical transition then requires two things to happen: first, that quantum features, like superposition and entanglement, must fade away; second, that an effective description of the macroscopic system must emerge from its quantum description. 
These issues become more prominent in a measuring process of a quantum system \cite{wheeler83,schlosshauer:1267,jacobs}. In such a situation, we interact a system, for instance a single atom, whose description is given by quantum mechanics, with a macroscopic measuring apparatus, for which a classical description is more suitable. The two realms, quantum and classical, meet in such situation. Much like in the Schrödinger's cat scenario, we do not expect to observe an entangled state between the atom and the macroscopic description of the apparatus. Also, for the measuring apparatus to be of any use, we should  observe the apparatus' pointer in a well defined position, and not on a superposition of possible classical values. That amounts for the apparatus to be described by its effective classical description. How are these traits obtained if we  depart from a fully quantum description? In other words, how do we reconcile our classical description of the measuring process with the fact that intrinsically both the system being measured and the measuring apparatus are mostly well described by quantum theory?

Traditionally, these questions are addressed by the formalism of decoherence \cite{Zurek2003,joos,Zurek1999}, which appeals to the unavoidable interaction between system+apparatus and the environment to explain the diminishing of quantum features. Nevertheless, this approach only cares for local observables and does not explain how the effective description of the apparatus emerges from its quantum mechanics many-body state. Without that, quantum properties of the apparatus could still be observed. 

The fact that a large quantum system might still have pronounced quantum features, if one has access to all its degrees of freedom, is nicely shown in references \cite{KoflerBrukner2007,KoflerBrukner2008}. In \cite{KoflerBrukner2007} the authors show that a large spin length still behaves in a quantum way if one can measure all the possible values of the spin, say, in the $z$-direction. It is only when the measurement outcomes are coarse-grained, i.e., when one cannot resolve nearby outcomes and integrate their signal, that a classical description is obtained. The approach devised in \cite{KoflerBrukner2007,KoflerBrukner2008}, however, is not dynamical, in the sense that it applies a coarse graining procedure directly on the measurement outcomes. Inspired by such results, in \cite{ourfirstart} the authors developed a framework that applies a coarse graining procedure directly on the many-body system. In this way they can obtain both the effective description of the system and its effective dynamics. The framework then combines the dynamical aspects of the decoherence approach with the coarse graining ideas of \cite{KoflerBrukner2007,KoflerBrukner2008}. 

The aim of the present article is thus to use the developed tools in \cite{ourfirstart} to analyze the effective dynamics of a quantum measurement process. In this way we hope to shed some light on one of the most intriguing points of the quantum formalism, the quantum measurement problem \cite{Ontheqmp2015}.  

Our article is organized as follows: we begin by explaining more rigorously the measurement process considered here and describing the coarse graining channel used, as well as its motivation. With the measurement scenario clear, we discuss: in section III how we can extract information about the populations of the density matrix of the system looking at the apparatus; in section IV the system-apparatus entanglement evolution in time (by evaluating the negativity) and the behavior of the quantum coherences of the reduced effective density matrix of the apparatus; and how these features changes when we make the coarse graining action stronger. In section V we present our conclusions.

\section{Modeling a quantum measurement process}
\label{sec:mqmp}

The global situation of the measurement process here modeled is similar to what is found in reference \cite{cathadav}: a system to be measured consisting of a qubit initially in a superposition $c_{0}\ket{0} + c_{1}\ket{1}$, $c_{0}$ and $c_{1} \in \mathbb{C}$, $|c_{0}|^2 + |c_{1}|^2 = 1$, and an apparatus consisting of $N >> 1$ qubits interact. The interaction Hamiltonian is such that it induces a conditioned rotation: depending on the state of the qubit-system the apparatus will rotate in one or the other direction in the spin coherent states space. Thus, information about $c_0$ and $c_1$ will be imprinted in the measuring device. Information which we recover measuring magnetization on the apparatus' effective state $\rho_{t}^{N}$, obtained via the action of a coarse graining channel $\Lambda_{\textrm{CG}}: \mathcal{L}(\mathcal{H}_{2^N}) \rightarrow \mathcal{L}(\mathcal{H}_{3})$ ($\mathcal{L}(\mathcal{H}_{i})$ is the space of linear operators acting on the Hilbert space $\mathcal{H}_i$), which reduces the dimension of the apparatus' Hilbert space. At the end of the process, we will have three possible outcomes in our measurement scenario, corresponding to three possible magnetization values: $1, 0$ and $-1$.

Specifically, the state of the qubit-system is given by $\ket{\psi_{0}} = c_0 \ket{0} + c_1 \ket{1}$, and $\ket{\Psi} = \sqrt{p}\ket{0} + \sqrt{1-p}e^{i \phi}\ket{1}$ is the state of each one of the $N$ apparatus' constituents -- $e^{i\phi}$ a phase and $p \in \, $[$0,1$]. The total initial state is then given by
\begin{equation}
\ket{\chi_0} = \ket{\psi_0}_{S} \otimes \ket{\Psi}^{N}_{A} = \ket{\psi_0}_{S} \otimes \big \{ \ket{\Psi}\otimes...\otimes\ket{\Psi} \big \}_{A}.
\label{initialtstate}
\end{equation}
The measuring device is initially in a product state, meaning that neither classical nor quantum correlations are present. In fact, neither between the apparatus' constituents nor between system to be measured and apparatus. The initial total system+apparatus density matrix, before any interaction, is given by the separable state
\begin{equation}
\begin{split}
    \proj{\chi_0} & = \, \big \{ |c_0|^2 \proj{0} + c_0 c_{1}^{*} | 0 \rangle\! \langle 1 |  \\
    & + c_1 c_{0}^{*} | 1 \rangle\! \langle 0 | + |c_1|^2 \proj{1} \big \} \otimes \proj{\Psi^{N}}.
\label{3initialtdm}
\end{split}
\end{equation}
The main motivation for such configuration came from systems of nuclear magnetic resonance (NMR), where the nuclear atomic spins play the role of qubits and  measures of magnetization and magnetic fields are of great importance \cite{NMR}. 

The interaction Hamiltonian, chosen by $H = \hbar \frac{\omega}{N} \sigma_{z} \otimes \vec{J_{x}}$,  is the responsible for generating entanglement between the parts. $\vec{J_x}$ is the vector sum of each individual angular momentum operator in the $x$-direction $\vec{J_{xi}}$, namely $\vec{J_x} = \vec{J_{x1}} + \vec{J_{x2}} +...+ \vec{J_{xN}}$, and $\omega$ the coupling constant, the angular frequency of the rotation induced by the angular momentum operator around the $x$-axis. 

The evolution will induce rotations in each apparatus' constituent, counterclockwise ($+\theta$) and clockwise ($-\theta$) directions, conditioned to the system's qubit states $\ket{0}$ and $\ket{1}$ ($\sigma_{z} \ket{0} = +1 \ket{0}$ and $\sigma_{z} \ket{1} = -1 \ket{1}$), respectively:
\begin{equation}
    \proj{\chi_t} = U_{t,0} \, \proj{\chi_0} U_{t,0}^{\dagger} = e^{-i \frac{\omega \, t}{N} \sigma_{z} \otimes \vec{J}_{x}} \proj{\chi_0} e^{i \frac{\omega \, t}{N} \sigma_{z} \otimes \vec{J}_{x}},
\label{initialtdm}
\end{equation}
since $e^{-i \frac{\omega \, t}{N} \vec{J}_{x}} = R_{\theta,x}$ and $\theta \equiv \frac{\omega t}{N}$. In equation \ref{initialtdm} we have a system+apparatus entangled state representing a rotation of $\theta$ or $-\theta$ (around the $x$-axis) in the apparatus' constituents conditioned to the eigenvalue $+1$ or $-1$ of the operator $\sigma_z$ acting on the system's state. 

To measure the system means to recover information about $c_0$ and $c_1$ looking at the measuring apparatus. Just as we are considering in our modeling the system inaccessible, we also do not have access to the measuring device in all its details, but via an effective description. Mathematically, this effective description will be provided by a coarse graining channel $\Lambda_{\textrm{CG}}$. The concept of coarse graining is the key aspect of this article. This is one way we translate mathematically the incapacity or the desire not to access completely the degrees of freedom of a given system. Or even intrinsic measurement errors. The decrease in the Hilbert space dimension represents the loss of partial information. However, the coarse graining operation is more general than simply tracing out some degrees of freedom of a given system \cite{ourfirstart}.

Notice that the output dimension of the coarse graining channel is $3$, which means that when measuring magnetization of the effective state of the apparatus (after the coarse graining action) in the $z$-direction three outcomes will be possible: $1$, $0$ and $-1$. Therefore, there will be some probability of finding $1, 0$ or $-1$ in the apparatus' display after a time $t$ of interaction, probabilities which will be related to the absolute values of $c_0$ and $c_1$. The analysis of such probabilities gives us intuition to construct $\Lambda_{\textrm{CG}}$.

\subsection{Constructing the coarse graining channel $\Lambda_{\textrm{CG}}$}

The apparatus' total magnetization is given by the sum $M_{z}=\sum_{i=1}^{N} \sigma_{z \, i}$,
where we are summing on all the $N$ constituents. This would be the case if we had access to each one of the $N$ qubits. However, we are considering the case in which we don't have such resolution. Inspired by the work done in \cite{MacroPoulin}, we decided to model this lack of resolution through a division in bins.

Our coarse description will be such that: if more than or $\frac{2}{3}$ of the $N$ apparatus' constituents are in the state $\ket{0}$, the measuring device will show in the display magnetization +1; if less than or $\frac{N}{3}$ are in the state $\ket{0}$, it will show magnetization -1; if in the interval $(\frac{N}{3},\frac{2N}{3})$, the apparatus will show magnetization 0. With this division in three bins, or regions [$0$,$\frac{N}{3}$], ($\frac{N}{3}$,$\frac{2N}{3}$) and [$\frac{2N}{3}$,$N$], projectors on different magnetization subspaces are being grouped and perceived as an effective one depending on the region that contains it. In other words, rather than perceiving changes in total magnetization with the resolution of one qubit, only changes involving $~\frac{N}{3}$ qubits are perceptible by the measuring device. In this context, the probability of obtaining magnetization $i$ will be represented by $\text{Pr}(Z_{i}^{N}| \Psi_t^{N})$, with $\Psi_{t}^{N} = \text{Tr}_{S}(\proj{\chi_t}) \in \mathcal{L}(\mathcal{H}_{2^{N}})$ the evolved reduced density matrix of the apparatus and $Z_{i}^{N}$ the sum of projectors on subspaces of the same total magnetization $i$ (for $i = 1$, see \ref{projectors}) as perceived by the not so good apparatus, representing sums of the POVM elements of the total magnetization on the z-direction and its permutations. For example, for $i = 1$, 
\begin{equation}
 Z_{1}^{N} = \sum_{l=\frac{2N}{3}}^{N} \sum_{\sigma} \Pi_{\sigma}| 0_{1} 0_{2} .. 0_{l} 1_{1} .. 1_{N-l} \rangle\! \langle 0_{1} 0_{2} .. 0_{l} 1_{1} .. 1_{N-l} | \Pi_{\sigma}^{\dagger}.
\label{projectors}
\end{equation}

As suggested by equation \ref{projectors}, the application of the coarse graining channel in the computational basis elements of a state of $N$ qubits, with $\Pi_{\sigma}$ the permutation operator, will be given by:
\begin{equation}
\begin{split}
\Lambda_{\textrm{CG}}(\Pi_\sigma | 0_{1}.. 0_{l} 1_{1} .. 1_{N-l} \rangle\! \langle 0_{1} .. 0_{l} 1_{1} .. 1_{N-l} | \Pi_{\sigma^{'}}^\dagger)& \\
= \, \Bigg\{
\begin{matrix}| 1 \rangle\!\langle 1 |,\,\,\, \mbox{if} \,\,\,  l \in [\frac{2N}{3},N] \,\,\, &\forall \, \sigma = \sigma^{'};  \\  
| 0 \rangle\!\langle 0 |,\,\,\, \mbox{if} \,\,\,  l \in (\frac{N}{3},\frac{2N}{3}) \,\,\, &\forall \,  \sigma = \sigma^{'}; \\
| -1 \rangle\!\langle -1 |,\,\,\, \mbox{if} \,\,\,  l \in [0,\frac{N}{3}] \,\,\, &\forall \, \sigma = \sigma^{'};  \end{matrix}
\end{split}
\label{cgattempt3n}
\end{equation}
and 0 $\forall \, \sigma \ne \sigma^{'}$. The states $\proj{1}$, $\proj{0}$ and $\proj{-1}$ represent the diagonal elements, from top to bottom, in our effective description $\in \mathcal{L}(\mathcal{H}_3)$. They are directly related with the magnetization outcomes $1$, $0$ and $-1$, respectively. The action of the channel on the elements $\Pi_\sigma | 0_{1} 0_{2} .. 0_{l} 1_{1} .. 1_{N-l} \rangle\! \langle 0_{1} 0_{2} .. 0_{l} 1_{1} .. 1_{N-l} | \Pi_{\sigma^{'}}^\dagger$ is $0$ by the fact that our coarse description does not distinguish between $\Pi_\sigma | 0_{1} 0_{2} .. 0_{l} 1_{1} .. 1_{N-l} \rangle$ and $\Pi_{\sigma^{'}} | 0_{1} 0_{2} .. 0_{l} 1_{1} .. 1_{N-l} \rangle$, two elements with the same number of $0$'s and $1$'s exchanged in a different way. Therefore, there can be no coherence between them.

In order to study the behavior of quantum coherences, we need to define the coarse graining channel action in the rest of the computational basis elements, the off-diagonal ones.

Consistently, for the coherence terms we define $\Lambda_{\textrm{CG}}( \Pi_\sigma | 0_{1} 0_{2} .. 0_{l} 1_{1} 1_{2} .. 1_{N-l} \rangle\! \langle 0_{1} 0_{2} .. 0_{l^{'}} 1_{1} 1_{2} .. 1_{N- l^{'}} | \Pi_{\sigma^{'}}^\dagger )$, $l \ne l'$, equal to:
\begin{equation}
\begin{split}
& n_{1}(N) \, | -1 \rangle\! \langle 1 |, \, \mbox{if} \,\,\,  l \in [0,\frac{N}{3}] \,\,\, \mbox{and} \,\,\,  l^{'} \in [\frac{2N}{3},N]; \\
& n_{2}(N) \, | -1 \rangle\! \langle 0 |, \, \mbox{if} \,\,\,  l \in [0,\frac{N}{3}] \,\,\, \mbox{and} \,\,\,  l^{'} \in (\frac{N}{3},\frac{2N}{3}); \\
& n_{3}(N) \, | 1 \rangle\! \langle 0 |, \, \mbox{if} \,\,\,   l \in [\frac{2N}{3},N] \,\,\, \mbox{and}  \,\,\, l^{'} \in (\frac{N}{3},\frac{2N}{3}); \\
& 0, \, \mbox{if} \,\,\, l \,\,\, \mbox{and} \,\,\,  l^{'}  \in [\frac{2N}{3},N], \,\,\, \mbox{or} \in [0,\frac{N}{3}], \,\,\, \mbox{or} \in (\frac{N}{3},\frac{2N}{3}). 
\end{split}
\label{cgattempt32n}
\end{equation}
The normalization factors $n_{i}(N)$ are given in the Appendix's first part (expression \ref{normfac}) and accounts for the number of elements in the computational basis that corresponds to the same effective element after the coarse graining action. Constructed the channel $\Lambda_{\textrm{CG}}$, whose complete positivity is shown in the Appendix's second part, we are now able to write the full effective state for the apparatus, $\rho_{t}^{N} = \text{Tr}_{S}[(\mathbb{1} \otimes \Lambda_{\textrm{CG}})(\proj{\chi_{t}})] \in \mathcal{L}(\mathcal{H}_{3})$ -- given in the Appendix's third part, expression \ref{effat}.

\section{Measuring the system looking at the apparatus' magnetization}
\label{sec:msla}

Defined the coarse graining channel and the measurement process considered, it is time to analyze its effects. Starting by looking at the probabilities $\text{Pr}_{i}(t)$ of measuring magnetization $i=1,0,-1$ in the effective state $\rho_{t}^{N}$, given by its populations. First, we should look at $\text{Pr}_{i}(0)$, just before the system+apparatus interaction. In other words, immediately after switching on the apparatus -- $t = 0$. 

\begin{figure*}[!th]
\includegraphics[scale=1.1]{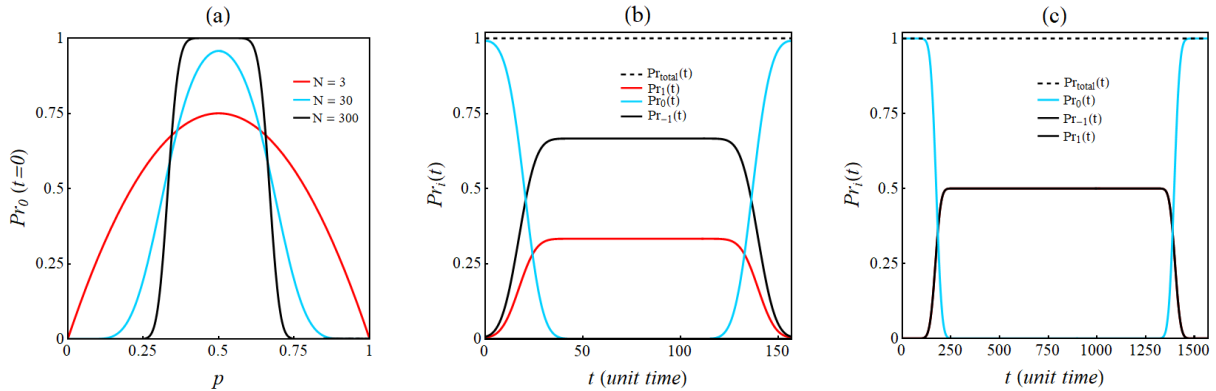} 
\caption{\small Probabilities of measuring magnetization: (a) $0$, as a function of the apparatus' constituents superposition coefficient $p$ before the interaction for three values of the number of apparatus constituents $N$. For large $N$ and $p$ around 0.5, the measuring device shows $0$ with probability $1$; (b) as a function of time for $N=50$ and system's state superposition coefficient $c_{0} =  \frac{1}{\sqrt{3}}$. The probabilities values for the plateaus are proportional to $|c_{0}|^{2}$ ($\textrm{Pr}_{1}(t)$) and $|c_{1}|^{2}$ ($\textrm{Pr}_{-1}(t)$); (c) as a function of time for $N=500$ and system's state superposition coefficient $c_{0} =  \frac{1}{\sqrt{2}}$. The probabilities values for the plateaus are proportional to $|c_{0}|^{2}$.}
\label{fig:Probmag}
\end{figure*}

Figure \ref{fig:Probmag} $(a)$ shows the probability of finding magnetization outcome $0$ in the apparatus display, before the interaction, as function of the superposition coefficient $p$ for the initial states of the apparatus' constituents, for three increasing values of the number of constituents $N$. Notice that, for large values of $N$, with the purpose of measuring magnetization in the z-axis, if we choose $p$ around $0.5$ the apparatus will show $0$ in the display. The probabilities of showing $1$ or $-1$ are zero. In fact, when turning on the measurement apparatus and before coupling it to the system, it is expected to show magnetization $0$. This result approaches our model to the real daily in the lab. The condition of a huge number of constituents is important to approach the situation with an apparatus with size that we are familiar with in daily-life.  

The graph shown in figure \ref{fig:Probmag} $(a)$ suggest the choice of $p = 0.5$. For simplicity, we have also chosen $\phi = \frac{\pi}{2}$ and $\omega =1$. In figures \ref{fig:Probmag} $(b)$ and \ref{fig:Probmag} $(c)$ are depicted plots of probabilities of obtaining outcome $1$,$0$ and $-1$ versus time for $N=50$ and $N=500$, respectively. For times greater than zero, the probabilities oscillate periodically, reaching local maximum whose values are given by $|c_0|^2$ ($\text{Pr}_{1}(t)$) and $|c_1|^2$ ($\text{Pr}_{-1}(t)$). Note that the oscillatory nature of the probabilities was expected, since the rotations to which the apparatus' constituents are subjected occur continuously.

The measurement of magnetization of the apparatus' effective state is with respect to the $z$-axis, with each of its $N$ constituents rotating over the $x$-axis from the initial state $\ket{\Psi} = \frac{1}{\sqrt{2}} \ket{0} + \frac{i}{\sqrt{2}}\ket{1}$; the $\theta$ angle grows positively in the counterclockwise direction. Considering the Bloch sphere for each constituent, it is initially in a plane which is perpendicular to the $z$-axis -- consider as ``the equator''. That is the reason for the apparatus to initially measure $0$. In fact, the probability of measuring $0$ is maximum when the constituents are close to the equator's plane in the Bloch sphere -- $\theta$ a multiple integer of $\pi$. As the interaction occurs, the apparatus remains in a superposition of rotating in two directions. The best moments to acquire information about the system being when the apparatus is close to the poles, aligned with the $z$-axis.

In the case of large values of $N$, interesting processes occur. First, the plateaus that are seen in the graphs \ref{fig:Probmag} $(b)$ and \ref{fig:Probmag} $(c)$ increase, meaning that for more and more time we have a well defined resulting outcome. I.e., the temporal region in which we can best acquire information about the system increase. Second, for large values of $N \, (N >> 1)$, the time required to go through a period will be infinite, since $\theta = \frac{\omega t}{N}$. In practice, in the limit $N \rightarrow \infty$ there will be no more oscillation. Thus, having waited for the time necessary to reach the plateau, the observer can look at the display in any time, being able to gain information about the system's initial state via $c_0$ and $c_1$. Again, our model approaches the daily-life situation.

\section{Decay of quantumness}
\label{sec:desa}

So far we have described a quantum measurement process in which by looking at an effective description of the measurement apparatus we recover information about a target system. With the help of figure \ref{fig:Probmag} we have studied the behavior of the probabilities and how to recover the superposition coefficients $c_0$ and $c_1$ of the target system. 

It is known that for an experimentalist who observes a classical (effective) description of an apparatus there is no quantum correlations between system and apparatus. In our model, therefore, we expect that in the limit of a huge number of constituents, or, in other words, in the limit of a strong coarse graining channel action, we can visualize the death of entanglement between system and apparatus. In order to visualize such effect, we numerically computed the bipartite Negativity $\mathcal{N}((\mathbb{1} \otimes \Lambda_{\textrm{CG}})(\proj{\chi_{t}}))$ -- a measure of entanglement given by $\mathcal{N}(\rho)=\text{Tr}\sqrt{(\rho^{\Gamma})^{\dagger}\rho^{\Gamma}}-1$, $\Gamma$ indicating partial transposition -- for increasing values of $N$. 

\begin{figure}[!th]
\includegraphics[width=0.7\linewidth]{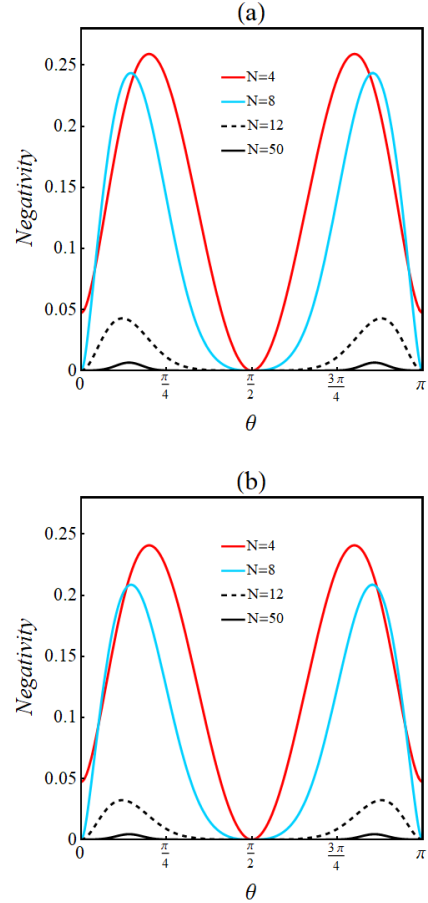}
\caption{\small Negativity for the total system+apparatus' effective state for increasing values of $N$. To increase the value of $N$ means also to increase the coarse graining action, resulting in a decrease in quantum correlations between system and measuring apparatus. The $\theta$ angle is in radians, $p = 0.5$, $\phi = \frac{\pi}{2}$ and $(a)$ $c_{0} = c_{1} = \frac{1}{\sqrt{2}}$ and $(b)$ $c_{0} = \frac{1}{\sqrt{3}}$.}
\label{negativity}
\end{figure}

We managed to see the death of entanglement as the number of apparatus' constituents $N$ grows, which means that the description is becoming more coarse. We exemplify this phenomenon in figure \ref{negativity}, with $N$ from $4$ to $50$ for the two cases already considered in the article, $(a)$ $c_0 = \frac{1}{\sqrt{2}}$ and $(b)$ $c_{0} = \frac{1}{\sqrt{3}}$. We should stress that due to numerical fluctuations near $\theta = 0$ and $\theta = n \pi$, $n$ odd, the negativity value for $N=4$ has a small difference to zero. However, the most important in figure $2$ is to observe that the negativity shows oscillatory and descending behavior with the increase in the apparatus' constituents number, becoming closer to zero as $N \rightarrow \infty$. Notice that we perceive the death of quantum correlations without evoking any interaction with the surrounding environment, but by the inability to completely resolve the internal degrees of freedom of the total system. 

Initially, target system and apparatus' effective state are almost not entangled. The evolution, characterized by a conditioned rotation, is the responsible for creating entanglement between the parts. After a rotation in the apparatus' state by an angle $\frac{\pi}{2}$, which represents an interaction time $t = \frac{N \pi}{2 \omega}$, the two possible states for the target system, that make up systems' superposition, are completely aligned with the $z$-axis. We have then a separable state. This is the best moment to measure apparatus magnetization and gather information about $c_{0}$ and $c_{1}$. Then the cycle starts over. 

\begin{figure*}[!th]
\includegraphics[scale=1.14]{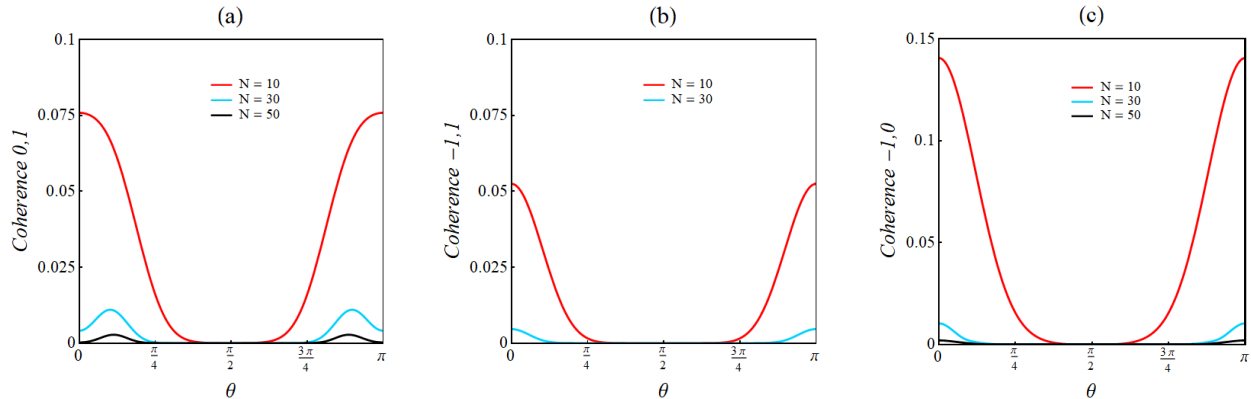} 
\caption{\small Decay of the absolute value of quantum coherences in the apparatus' reduced effective state. Each graph corresponds to one coefficient of the matrix representing the  effective reduced state of the apparatus $\rho_{t}^{N}$. The greater the number $N$ of the apparatus constituents, the lower the absolute values of the coherences. The $\theta$ angle is in radians, $p = 0.5$, $\phi = \frac{\pi}{2}$ and $c_{0} = \frac{1}{\sqrt{2}}$.}
\label{CoherModel3}
\end{figure*}

We were able to visualize the decay of quantum correlations between system and apparatus, but it would also be very interesting if the reduced effective density matrix of the apparatus $\rho_{t}^{N}$ -- given in Appendix's third part --  lost the off-diagonal terms in the limit of a strong coarse graining action. We would then have a density matrix representing a statistical mixture of possible states, only with the diagonal terms different from zero, corresponding to magnetization $1$, $0$ and $-1$. In fact, it happens in our model. To exemplify, the graphs with the behavior of the coherences in the effective reduced state of the apparatus $\rho_{t}^{N}$ for $c_{0} = \frac{1}{\sqrt{2}}$ are shown in figure \ref{CoherModel3}.   

\section{Conclusions}

We presented a model of a quantum measurement process in which the target system is a two level system and the measuring device composed by N two level systems. One of the goals achieved was to recover information about the coefficients $c_0$ and $c_1$ of the system. For such, the system+apparatus dynamics corresponded to a rotation in the apparatus conditioned to the system's state $\ket{0}$ and $\ket{1}$. Using the constructed coarse graining channel allowed us to model a digital measurement scenario with three outcomes, representing magnetization positive, negative and zero. An interesting fact is that the outcome zero appears when we turn on the measuring device, in the limit of a large apparatus. This is a situation closer to everyday life in the lab. We also visualize: the death of quantum correlations between system and apparatus; the death of quantum coherences in the apparatus' reduced effective state the coarser the description. In particular, for large $N$ the effective state $\rho_{t}^{N}$ can be seen as a statistical mixture of possibilities.

This is one of the main results obtained by the formalism of decoherence. We get the same using a mathematical formalism which is more general, since applying a coarse graining channel doesn't just mean a partial trace \cite{ourfirstart}. It wasn't necessary either to evoke degrees of freedom external to system+apparatus, for instance any interaction with the surrounding environment. In Schrödinger's cat context, the cat can be perceived as a pretty rough description of the apparatus and the poison as the system. The fact that we do not perceive the cat in its most detailed quantum description would be the responsible for the disappearance of quantum features.

The main message is that our not so good, blurry description of the world may have central hole in the fact that we did not observe quantumness in our daily lives. Our work confirms our intuition of the death of quantum correlations due to the lack of access to all system's degrees of freedom and shed some light on a way to attack the quantum measurement problem \cite{Ontheqmp2015}.


\begin{acknowledgments}
Special thanks goes to Dr. Fernando de Melo and Dr. \v{C}aslav Brukner for stimulating discussions, and the referee for insightful observations and valuable feedback. This work is supported by the Brazilian funding agencies CNPq and CAPES, and it is part of the Brazilian National Institute for Quantum Information.
\end{acknowledgments}



\appendix

\begin{widetext}
\section{}
\label{appendix}
\textbf{First part: normalization factors.} The expressions for the normalization factors $n_{1}(N)$, $n_{2}(N)$ and $n_{3}(N)$ in equation \ref{cgattempt32n} are given by
\begin{equation}
\begin{split}
n_{1}(N) & = \bigg \{ \sum_{k=0}^{\frac{N}{3}} \sum_{n=\frac{2N}{3} - k}^{N-k} \sum_{m=\frac{2N}{3}-n}^{N-n-k} \frac{N!}{k!n!m!(N-n-k-m)!}  \bigg \}^\frac{-1}{2}; \\
n_{2}(N) & = \bigg \{ \sum_{k > \frac{N}{3}}^{< \frac{2N}{3}} \sum_{n=\frac{2N}{3}-k}^{N-k} \sum_{m>\frac{N}{3}-k}^{<\frac{2N}{3}- k} \frac{N!}{k!n!m!(N-n-k-m)!}  \bigg \}^\frac{-1}{2}; \\
n_{3}(N) & = \bigg \{ \sum_{k > \frac{N}{3}}^{< \frac{2N}{3}} \sum_{n > \frac{N}{3} - k}^{<\frac{2N}{3}-k} \sum_{m=0}^{\frac{N}{3} - n} \frac{N!}{k!n!m!(N-n-k-m)!}  \bigg \}^\frac{-1}{2}.
\label{normfac}
\end{split}    
\end{equation}

\textbf{Second part: complete positivity of $\Lambda_{\textrm{CG}}$.} In order to show that the coarse graining channel $\Lambda_{\textrm{CG}}$ is completely positive, we must show that $\bra{v}\!\rho_{\textrm{CG}}\!\ket{v} \ge 0 \,\,\,\, \forall \, \ket{v} \, \in \, \mathcal{H}_{2^N.3}$, with $\rho_{\Lambda_{\textrm{CG}}}$ the corresponding Choi matrix of the channel $\Lambda_{\textrm{CG}}$ (equation \ref{choimatrix}). 
\begin{equation}
\rho_{\Lambda_{\textrm{CG}}} = (\idty \otimes \Lambda_{\textrm{CG}})(\proj{\Phi^{+}})\textrm{,} \,\,\, \textrm{with} \,\,\,  \ket{\Phi^{+}} = \frac{1}{\sqrt{2^N}} \sum_{i_{1}=0}^{1} \sum_{i_{2}=0}^{1} . . . \sum_{i_{N}=0}^{1} \ket{i_{1} i_{2} ... i_{N}} \otimes \ket{i_{1} i_{2} ... i_{N}}.
\label{choimatrix}
\end{equation}
Writing $\ket{v} = \sum_{\vec{k}, l} c_{k_{1} k_{2} ... k_{N}, l} \ket{k_{1} k_{2} ... k_{N}, l}$ and $\bra{v} = \sum_{\vec{k}, l} c_{k_{1} k_{2} ... k_{N}, l}^{*} \bra{k_{1} k_{2} ... k_{N}, l}$, with each element of $\vec{k}$, $k_{i}$, $\in \{0,1\}$, $l \in \{1,0,-1\}$ and $\sum_{\vec{k}, l} |c_{k_{1} k_{2} ... k_{N} ,l}|^2 = 1$,
\begin{equation}
\bra{v}\!\rho_{\textrm{CG}}\!\ket{v} = \sum_{\vec{k}, l} c_{k_{1} k_{2} ... k_{N}, l}^{*} \bra{k_{1} k_{2} ... k_{N}, l} \sum_{\vec{i}, \vec{j}} \ket{i_{1} i_{2} ... i_{N}}\!\bra{j_{1} j_{2} ... j_{N}} \otimes \Lambda_{\textrm{CG}}(\ket{i_{1} i_{2} ... i_{N}}\!\bra{j_{1} j_{2} ... j_{N}}) \frac{1}{2^N} \sum_{\vec{p}, q} c_{p_{1} p_{2} ... p_{N}, q} \ket{p_{1} p_{2} ... p_{N}, q}.
\end{equation}
Acting the coarse graining channel, 
\begin{equation}
\begin{split}
\bra{v}\!\rho_{\textrm{CG}}\!\ket{v}&= \sum_{\vec{k}, l, \vec{i}, \vec{j}, \vec{p}, q, m, n} c_{k_{1} ...,l}^{*} c_{p_{1} ... ,q} \bra{k_{1} k_{2} ... k_{N}, l}\!\ket{i_{1} i_{2} ... i_{N}, m}\!\bra{j_{1} j_{2} ... j_{N}, n}\!\ket{p_{1} p_{2} ... p_{N}, q} \frac{\mathcal{N}(N,m,n)}{2^N}\\
\Rightarrow \bra{v}\!\rho_{\textrm{CG}}\!\ket{v} &= \sum_{\vec{i},\vec{j}, m, n}  c_{i_{1} ...,m}^{*} c_{j_{1} ... ,n} \frac{\mathcal{N}(N,m,n)}{2^N},
\end{split}
\end{equation}
with each element of $\vec{i}$, $i_{i}$, and each element of $\vec{j}$, $j_{i}$, $\in \{0,1\}$, $m$ and $n \in \{1,0,-1\}$ and $\mathcal{N}(N,m,n) \in \{1, n_{1}(N), n_{2}(N), n_{3}(N)\}$, assuming the value $1$ when $m=n$. 

We can rewrite $\bra{v}\!\rho_{\textrm{CG}}\!\ket{v}$ as 
\begin{equation}
\bra{v}\!\rho_{\textrm{CG}}\!\ket{v}=\frac{1}{2^N} \{ \sum_{\vec{i} = \vec{j}} |c_{i_{1} ... ,1}|^{2} + \sum_{\vec{i} = \vec{j}} |c_{i_{1} ... ,0}|^{2} + \sum_{\vec{i} = \vec{j}} |c_{i_{1} ... ,-1}|^{2} \} + \sum_{\vec{i},\vec{j}, m \ne n} c_{i_{1} ... ,m}^{*} c_{j_{1} ... ,n} \frac{\mathcal{N}(N,m,n)}{2^N}.
\end{equation} 
Since we are considering the case where the vector $\ket{v}$ is normalized, we demand that, $\forall \ket{v}$ and $\forall N$,
\begin{equation}
1 + \sum_{\vec{i},\vec{j}, m \ne n} c_{i_{1} ... ,m}^{*} c_{j_{1} ... ,n} \mathcal{N}(N,m,n) \ge 0,
\end{equation}
or,
\begin{equation}
\begin{split}
 1 & +  \sum_{\vec{i},\vec{j}} c_{i_{1} ... -1}^{*} c_{j_{1} ... 1} \, n_{1}(N) + \sum_{\vec{i},\vec{j}} c_{i_{1} ... -1}^{*} c_{j_{1} ... 0} \, n_{2}(N) + \sum_{\vec{i},\vec{j}} c_{i_{1} ... 0}^{*} c_{j_{1} ... 1} \, n_{3}(N) \\ 
& +  \sum_{\vec{i},\vec{j}} c_{i_{1} ... 1}^{*} c_{j_{1} ... -1} \, n_{1}(N)  + \sum_{\vec{i},\vec{j}} c_{i_{1} ... 0}^{*} c_{j_{1} ... -1} \, n_{2}(N) + \sum_{\vec{i},\vec{j}} c_{i_{1} ... 1}^{*} c_{j_{1} ... 0} \, n_{3}(N)   \ge 0.
\label{sums}
\end{split}
\end{equation}
The products of $c$ are the off-diagonal elements in the density matrix $\proj{v}$.The possible values $i_{i}$ and $j_{i}$ can assume in each sum above are conditioned to the values of $m$ and $n$, respecting the construction of the coarse graining channel. For instance, in the first sum, where $m = -1$ and $n = 1$, the number of $0$'s in the sequence $i_{1} ... i_{N}$  must be less or equal than $\frac{N}{3}$ and the number of $0$'s in the sequence $j_{1} ... j_{N}$ must be greater than or equal to $\frac{2N}{3}$. It also means that the number of terms in each sum in \ref{sums} is known.

Notice that 
\begin{equation}
\sum_{\vec{i},\vec{j}} c_{i_{1} ... ,-1}^{*} c_{j_{1} ... ,1} + \sum_{\vec{i},\vec{j}} c_{i_{1} ... ,1}^{*} c_{j_{1} ... ,-1} = \sum_{\vec{i},\vec{j}} 2 \,\, Re[c_{i_{1} ... ,-1}^{*} c_{j_{1} ... ,1}].
\end{equation}
And that
\begin{equation}
\sum_{\vec{i},\vec{j}} 2 \,\, Re[c_{i_{1} ... ,-1}^{*} c_{j_{1} ... ,1}] \ge -2\sum_{\vec{i},\vec{j}} \,\,| Re[c_{i_{1} ... ,-1}^{*} c_{j_{1} ... ,1}]|. 
\end{equation}
Therefore, we demand that $\forall \ket{v}$ and $\forall N$, 
\begin{equation}
\begin{split}
1 -  2 \, n_{\textrm{mín}}(N) \, \{ \sum_{\vec{i},\vec{j}} |Re[c_{i_{1} ... ,-1}^{*} c_{j_{1} ... ,1}]| +  \sum_{\vec{i},\vec{j}} |Re[c_{i_{1} ... ,-1}^{*} c_{j_{1} ... ,0}]| +  \sum_{\vec{i},\vec{j}} |Re[c_{i_{1} ... ,0}^{*} c_{j_{1} ... ,1}]| \} \ge 0,
\end{split}
\label{last}
\end{equation}
where we substitute $n_{1}(N)$, $n_{2}(N)$ and $n_{3}(N)$ by the smallest of them.
 
Since $\sum_{\vec{k}, l} |c_{k_{1} k_{2} ... k_{N} ,l}|^2 = 1$ and its number of terms is equal to $3. 2^{N}$, the sum of modulus in \ref{last} will have no greater growth with $N$ than if we substitute each term in each sum by $\frac{5}{3.2^{N}}$. Making the replacement,  expression \ref{last} becomes $1 - f(N) \ge 0$. With $f(N)$ a known function of $N$, since we have expressions \ref{normfac} and the number of terms in each \ref{last} sums, which are $n_{1}(N)^{-2}$, $n_{2}(N)^{-2}$ and $n_{3}(N)^{-2}$, respectively. Then, it can be seen that indeed $f(N) \le 1$ $\forall N$. These make our demand \ref{sums} true.

\textbf{Third part: apparatus' effective state.} Consider the apparatus' effective reduced density matrix $\rho_{t}^{N} = \text{Tr}_{S}[(\mathbb{1} \otimes \Lambda_{\textrm{CG}})(\proj{\chi_{t}})]$, given by
\begin{equation}
\rho_{t}^{N} = |c_0|^2  \, \Lambda_{\textrm{CG}}(R_{\theta,x} \, \proj{\Psi^{N}} R_{\theta,x}^{\dagger}) + |c_1|^2 \, \Lambda_{\textrm{CG}}(R_{-\theta,x} \, \proj{\Psi^{N}} R_{-\theta,x}^{\dagger}).
\label{fulleffeaftercg}
\end{equation}
The evolved total state $R_{\theta,x} \, \proj{\Psi^{N}} R_{\theta,x}^{\dagger} = e^{-i \frac{\omega \, t}{N} \sigma_{z} \otimes \vec{J}_{x_{1}}} \proj{\Psi} e^{i \frac{\omega \, t}{N} \sigma_{z} \otimes \vec{J}_{x_{1}}} \otimes ... \otimes e^{-i \frac{\omega \, t}{N} \sigma_{z} \otimes \vec{J}_{x_{N}}} \proj{\Psi} e^{i \frac{\omega \, t}{N} \sigma_{z} \otimes \vec{J}_{x_{N}}}$  can be written as
\begin{equation}
    \left[  {\begin{array}{cc} x & x_{c}^{*} \\ x_{c} & 1-x \end{array} } \right]_{1} \otimes  \left[  {\begin{array}{cc} x & x_{c}^{*} \\ x_{c} & 1-x \end{array} } \right]_{2} \otimes
    .. \otimes
    \left[  {\begin{array}{cc} x & x_{c}^{*} \\ x_{c} & 1-x \end{array} } \right]_{N},
\label{totalatensorproduct}    
\end{equation}
with subscript $c$ meaning coherence, $x \equiv \big\{ \frac{1}{2} + (-\frac{1}{2} +p)\cos{\theta} + \sqrt{p(1-p)}\sin{\theta}\sin{\phi} \big\}$ and $x_c \equiv \frac{1}{2} \big\{i(1-2p)\sin{\theta} + 2\sqrt{p-p^2}(\cos{\phi} + i \cos{\theta}\sin{\phi}) \big\}$. Finding the general $2^{N} \times 2^{N}$ matrix resulting from \ref{totalatensorproduct}, applying the coarse graining channel in its elements and inserting the result in \ref{fulleffeaftercg}, not forgetting the contribution coming from $|c_1|^2 \, \Lambda_{\textrm{CG}}(R_{-\theta,x} \, \proj{\Psi^{N}} R_{-\theta,x}^{\dagger})$, is it possible to get $\rho_{t}^{N}$ -- we get $y$ from $x$ doing $\theta \rightarrow -\theta$: 

\begin{equation}
\begin{split}
\rho_{t}^{N} & = \proj{1} \Big( |c_0|^2 \sum_{k=\frac{2N}{3}}^{N} \binom{N}{k} x^{k} (1-x)^{N-k} + |c_1|^2 \sum_{k=\frac{2N}{3}}^{N} \binom{N}{k} y^{k} (1-y)^{N-k} \Big) \\
& + \proj{0} \Big( |c_0|^2 \sum_{k > \frac{N}{3}}^{< \frac{2N}{3}} \binom{N}{k} x^{k} (1-x)^{N-k} + (1-|c_0|^2) \sum_{k > \frac{N}{3}}^{< \frac{2N}{3}} \binom{N}{k} y^{k} (1-y)^{N-k} \Big) \\
& + \proj{-1} \Big( |c_0|^2 \sum_{k=0}^{\frac{N}{3}} \binom{N}{k} x^{k} (1-x)^{N-k} + (1-|c_0|^2) \sum_{k=0}^{\frac{N}{3}} \binom{N}{k} y^{k} (1-y)^{N-k} \Big) \\
& + | 1 \rangle\! \langle 0 | \, n_{3}(N) \, \Big( |c_0|^2 \sum_{k>\frac{N}{3}}^{<\frac{2N}{3}} x^k \sum_{n>\frac{N}{3} - k}^{<\frac{2N}{3}-k} x_{c}^{n} \sum_{m=0}^{\frac{N}{3}-n} (1-x)^{m} (x_{c}^{*})^{N-k-n-m} \frac{N!}{k!n!m!(N-n-k-m)!} \\
& + |c_1|^2 \sum_{k>\frac{N}{3}}^{<\frac{2N}{3}} y^k \sum_{n>\frac{N}{3} - k}^{<\frac{2N}{3}-k} y_{c}^{n} \sum_{m=0}^{\frac{N}{3}-n} (1-y)^{m} (y_{c}^{*})^{N-k-n-m} \frac{N!}{k!n!m!(N-n-k-m)!} \Big) \\
& + | 1 \rangle\! \langle -1 | \, n_{1}(N) \, \Big( |c_0|^2 \sum_{k=0}^{\frac{N}{3}} x^k \sum_{n=\frac{2N}{3} - k}^{N-k} x_{c}^{n} \sum_{m=\frac{2N}{3}-n}^{N-n-k} (1-x)^{m} (x_{c}^{*})^{N-k-n-m} \frac{N!}{k!n!m!(N-n-k-m)!} \\
& + |c_1|^2 \sum_{k=0}^{\frac{N}{3}} y^k \sum_{n=\frac{2N}{3} - k}^{N-k} y_{c}^{n} \sum_{m=\frac{2N}{3}-n}^{N-n-k} (1-y)^{m} (y_{c}^{*})^{N-k-n-m} \frac{N!}{k!n!m!(N-n-k-m)!} \Big) \\
& + | -1 \rangle\! \langle 0 | \, n_{2}(N) \, \Big( |c_0|^2 \sum_{k>\frac{N}{3}}^{<\frac{2N}{3}} x_{c}^k \sum_{n=\frac{2N}{3} - k}^{N-k} (1-x)^n \sum_{m>\frac{N}{3}-k}^{<\frac{2N}{3}-k} x^{m} (x_{c}^{*})^{N-k-n-m} \frac{N!}{k!n!m!(N-n-k-m)!} \\
& + |c_1|^2 \sum_{k>\frac{N}{3}}^{<\frac{2N}{3}} y_{c}^k \sum_{n=\frac{2N}{3} - k}^{N-k} (1-y)^n \sum_{m>\frac{N}{3}-k}^{<\frac{2N}{3}-k} y^{m} (y_{c}^{*})^{N-k-n-m} \frac{N!}{k!n!m!(N-n-k-m)!}  \Big) \\
& + | 0 \rangle\! \langle -1 | \, n_{2}(N) \, \Big( |c_0|^2 \sum_{k>\frac{N}{3}}^{<\frac{2N}{3}} (x_{c}^{*})^k \sum_{n=\frac{2N}{3} - k}^{N-k} (1-x)^n \sum_{m>\frac{N}{3}-k}^{<\frac{2N}{3}-k} x^{m} x_{c}^{N-k-n-m} \frac{N!}{k!n!m!(N-n-k-m)!} \\
& + |c_1|^2 \sum_{k>\frac{N}{3}}^{<\frac{2N}{3}} (y_{c}^{*})^k \sum_{n=\frac{2N}{3} - k}^{N-k} (1-y)^n \sum_{m>\frac{N}{3}-k}^{<\frac{2N}{3}-k} y^{m} y_{c}^{N-k-n-m} \frac{N!}{k!n!m!(N-n-k-m)!}  \Big) \\
& + | -1 \rangle\! \langle 1 | \, n_{1}(N) \, \Big( |c_0|^2 \sum_{k=0}^{\frac{N}{3}} x^k \sum_{n=\frac{2N}{3} - k}^{N-k} (x_{c}^{*})^n \sum_{m=\frac{2N}{3}-n}^{N-n-k} (1-x)^{m} x_{c}^{N-k-n-m} \frac{N!}{k!n!m!(N-n-k-m)!} \\
& + |c_1|^2 \sum_{k=0}^{\frac{N}{3}} y^k \sum_{n=\frac{2N}{3} - k}^{N-k} (y_{c}^{*})^n \sum_{m=\frac{2N}{3}-n}^{N-n-k} (1-y)^{m} y_{c}^{N-k-n-m} \frac{N!}{k!n!m!(N-n-k-m)!} \Big)  \\
& + | 0 \rangle\! \langle 1 | \, n_{3}(N) \, \Big( |c_0|^2 \sum_{k>\frac{N}{3}}^{<\frac{2N}{3}} x^k \sum_{n>\frac{N}{3} - k}^{<\frac{2N}{3}-k} (x_{c}^{*})^{n} \sum_{m=0}^{\frac{N}{3}-n} (1-x)^{m} x_{c}^{N-k-n-m} \frac{N!}{k!n!m!(N-n-k-m)!} \\
& + |c_1|^2 \sum_{k>\frac{N}{3}}^{<\frac{2N}{3}} y^k \sum_{n>\frac{N}{3} - k}^{<\frac{2N}{3}-k} (y_{c}^{*})^{n} \sum_{m=0}^{\frac{N}{3}-n} (1-y)^{m} y_{c}^{N-k-n-m} \frac{N!}{k!n!m!(N-n-k-m)!} \Big).
\label{effat}
\end{split}
\end{equation}

In equation \ref{effat}, notice that each coherence term involves, in its numerator and denominator, equal sums containing the factor $\frac{N!}{k!n!m!(N-n-k-m)!}$, all sums growing with $N$. In numerator are also present quantities in absolute values smaller than $1$ -- they are $x$, $x_{c}$, $y$, $(1-x)$, $(1-y)$ and $y_{c}$ -- to powers which, summed up, are of the order of $N$. This ensures that for large values of $N$, i. e. in the limit of a huge measuring apparatus, for instance consisting of moles of atoms, the coherences vanishes. Not only by the size of the apparatus, but mainly by our coarser description, the inability to access it in all its degrees of freedom. 

\end{widetext}

\end{document}